\documentstyle [epsf]{article}
\begin{document}

\title{\bf Parametrically Excited Surface Waves: Two-Frequency
Forcing, Normal Form Symmetries, and Pattern Selection}

\author{Mary Silber\\ 
{\it Department of Engineering Sciences \& Applied Mathematics}\\ 
{\it Northwestern University}\\ 
{\it Evanston, IL 60208, USA} 
\and{Anne C. Skeldon}\\ 
{\it Department of Mathematics}\\ 
{\it City University, Northampton Square}\\ 
{\it London, EC1V OHB, UK}}

\date{October 1, 1998}

\maketitle
\begin{abstract}
Motivated by experimental observations of exotic free surface standing
wave patterns in the two-frequency Faraday experiment, we investigate
the role of normal form symmetries in the associated pattern selection
problem.  With forcing frequency components in ratio $m/n$, where
$m$ and $n$ are co-prime integers that are not both odd, there is the
possibility that both harmonic waves and subharmonic waves may lose
stability simultaneously, each with a different wavenumber. We focus
on this situation and compare the case where the harmonic waves have a
longer wavelength than the subharmonic waves with the case where the
harmonic waves have a shorter wavelength. We show that in the former
case a normal form transformation can be used to remove all
quadratic terms from the amplitude equations governing the relevant
resonant triad interactions. Thus the role of resonant triads in the
pattern selection problem is greatly diminished in this situation. We
verify our general bifurcation theoretic results within the example of
one-dimensional surface wave solutions of the Zhang-Vi\~{n}als
model~\cite{ref:zv97} of the two-frequency Faraday problem. In
one-dimension, a $1:2$ spatial resonance takes the place of a resonant
triad in our investigation. We find that when the bifurcating modes
are in this spatial resonance, it dramatically effects the bifurcation
to subharmonic waves in the case that the forcing frequencies are in
ratio $1/2$; this is consistent with the results
in~\cite{ref:zv97}. In sharp contrast, we find that when the forcing
frequencies are in ratio $2/3$, the bifurcation to (sub)harmonic waves
is insensitive to the presence of another spatially-resonant
bifurcating mode. This is consistent with the results of our general
analysis.

\end{abstract}

PACS numbers: 47.54.+r, 47.20.Ky, 47.35.+i

\section{Introduction.}
\label{sec:intro}

Exotic free surface standing wave patterns, parametrically excited by
two-frequency forcing, have attracted considerable attention in recent
years, both experimentally \cite{ef:93,ef:94,ref:muller93,gollub} and
theoretically
\cite{ref:zv97,friedrich,laurette,ref:lp97,sp}. 
In this system, the surface waves are
excited by subjecting the fluid layer to a time-periodic vertical
acceleration with two (rationally-related) frequency components. This
corresponds to a modification of a classic hydrodynamic problem, which
dates back to observations of Faraday \cite{farad}, in which the
surface waves are parametrically excited by a purely sinusoidal
vertical acceleration of the fluid container
\cite{miles}. 
Triangular patterns \cite{ref:muller93}, quasi-patterns
\cite{ef:94,gollub}, and superlattice patterns \cite{gollub} are among the more
exotic states that have been observed in laboratory experiments employing
two-frequency forcing.  Here, ``quasi-patterns'' are patterns with twelve-fold
symmetry and  are the hydrodynamic analogues of
two-dimensional quasi-crystals. These patterns are not spatially-periodic, but
do have long-range orientational order; their spatial Fourier transform
exhibits twelve prominent equally-spaced peaks that lie on a circle.  Although
such patterns have been observed in single-frequency Faraday experiments
\cite{cal:92,cal:95,binks}, 
they are more readily observed with two-frequency forcing 
and examples have been seen with excitation frequencies in ratio 4/5,
4/7, 6/7, 8/9 \cite{ef:94}.  In contrast, ``superlattice'' patterns
are spatially-periodic with structure on two disparate
length-scales. They have been observed in the two-frequency Faraday
system with a forcing frequency ratio of 6/7~\cite{gollub}.  The
observations of both quasipatterns and superlattice patterns have
indicated that the patterns are {\it synchronous} with the forcing
frequency.  The Faraday wave problem is an attractive experimental
system for studying the fundamental mechanisms behind the occurrence
of such patterns because of the fast time scales involved and the
number of tunable control parameters
\cite{bechhoefer,kg96}. 

One motivation behind some of the early experiments on two-frequency
parametric excitation of Faraday waves was to destroy the $Z_2$ symmetry
associated with subharmonic waves
\cite{ef:93,ef:94,ref:muller93}. Specifically, in the classic Faraday
experiment with single-frequency forcing $\cos(\omega t)$, the onset
surface waves respond subharmonically with frequency $\omega/2$
\cite{kt}. (Harmonic response is also possible for very shallow layers of
viscous fluid~\cite{kumar96,muller97}.) In the situation of subharmonic
response there is a discrete time translation symmetry $t\to
t+{2\pi\over\omega}$ that is broken by the state of the system. This
symmetry-breaking manifests itself in the governing pattern amplitude
equations by suppressing all even terms. Whether or not this extra
$Z_2$ symmetry is present has a particularly profound effect on the
formation of hexagonal patterns \cite{gsk}; for instance, in many
systems where it is absent, hexagonal patterns are preferred at
onset~\cite{cross}. Edwards and Fauve \cite{ef:94} noted that by
introducing the second perturbing frequency component to the periodic
forcing, they could destroy the discrete time translation symmetry of
the system.

A second feature of using two frequency forcing is that it is possible
to obtain a neutral stability curve with minima at two distinct
wavenumbers, where the ratio of these two critical wavenumbers can be
adjusted by varying the two frequency components of the
forcing. Typically, one minimum corresponds to standing waves that are
subharmonic with respect to the forcing frequency, while the other
minimum corresponds to synchronous waves \cite{laurette}.  It was
proposed by Edwards and Fauve~\cite{ef:94} that by tuning the ratio of
wavenumbers of the most unstable modes one could control the resonant
triad interactions that are important to the pattern formation
process~\cite{newell}. Indeed all of the exotic patterns mentioned
above were obtained with experimental parameters near the so-called
``bicritical point'' where two modes lose stability simultaneously
\cite{ef:94, ref:muller93, gollub}.

The role of resonant triad interactions in the formation of Faraday wave
patterns has been investigated extensively by Vi\~{n}als and co-workers for
both the case of single frequency forcing~\cite{ref:cv97,ref:zv97a}, and
two-frequency forcing~\cite{ref:zv97}. The most detailed two-frequency
calculations focused on the situation where the frequency ratio was 1/2, and
the onset surface wave response was subharmonic with the forcing. Zhang and
Vi\~{n}als compared their theoretical results with experimental results of
M\"uller~\cite{ref:muller93}, who observed subharmonic hexagons, triangles and
squares near the bicritical point; which pattern was observed depended on a
relative phase between the $\omega$ and $2\omega$ sinusoidal waveforms in the
forcing function. A key theoretical idea behind the work of Zhang and
Vi\~{n}als~\cite{ref:zv97,ref:zv97a}, and Chen and Vi\~{n}als~\cite{ref:cv97}
is that the presence of certain resonant triads composed of Fourier mode
wavevectors ${\bf k}_1$, ${\bf k}_2$ and ${\bf k}_1+{\bf k}_2$ ($|{\bf
k}_1|=|{\bf k}_2|$) can suppress the formation of regular wave patterns that
involve the ${\bf k}_1$ and ${\bf k}_2$ modes. In particular, this is the case
when the ${\bf k}_1$ and ${\bf k}_2$ modes are excited at the onset of the
Faraday wave instability, while the mode with wave vector ${\bf k}_1+{\bf k}_2$
is only weakly damped. This situation arises quite naturally near the
bicritical point in parameter space.

Motivated by the ubiquity of quasi-patterns in two-frequency Faraday
experiments, Lifshitz and Petrich~\cite{ref:lp97} recently
investigated twelve-fold quasi-patterns within the framework of a
simple Swift-Hohenberg type model for the evolution of a real scalar
field $u({\bf x},t)$. Their model lacks reflection-symmetry $u\to -u$,
and the linearized equations lead to a neutral stability curve with a
double absolute minimum. However, because their model leads to
time-independent patterns arising from steady state bifurcation, it
does not capture one of the key features of the two-frequency Faraday
problem. Specifically, at the bicritical point in the Faraday problem,
one of the neutral curve minima is associated with a Floquet
multiplier (FM) $+1$, while the other minimum is associated with a FM
of $-1$.  In this paper we are specifically interested in this feature
of the bicritical point of the two-frequency Faraday experiment, and
its potential influence on the resonant triad interactions important
to the pattern formation problem.

We investigate the role of resonant triads in the formation of wave
patterns near the bicritical point in parameter space.  We focus on
the situation that applies to the two-frequency
quasipattern~\cite{ef:94} and superlattice pattern~\cite{gollub}
experiments in which the forcing frequency ratio, $m/n<1$, has $m$
even and $n$ odd. We show that the usual contribution of resonant
triads to the cross-coupling coefficient in the amplitude equations
may be greatly {\it suppressed} in this case. This is in marked
contrast to the situation of subharmonic waves with forcing ratio
$m/n<1$, where $m$ is odd and $n$ is even, {\it e.g.}  $m/n=1/2$, as
studied by Zhang and Vi\~{n}als. In this latter case, resonant triads
lead to peaks in the cross-coupling coefficient
$g(\theta)$~\cite{ref:zv97}.  In order to understand the distinction
between these two situations, we first recall that in each of these
cases the neutral stability curve has a double absolute minimum at the
bicritical point, with one of the minima associated with a FM$=+1$ and
the other with a FM$=-1$~\cite{ef:94,laurette}. At the bicritical
point, there are two distinct critical wavenumbers, $k_m$ and $k_n$
($k_m<k_n$). The FM$=+1$ is associated with the smaller critical
wavenumber $k_m$ if the even frequency is less than the odd one ({\it
i.e.}, $m/n={\rm even/odd}$); and it is associated with the larger
wavenumber $k_n$ if the even frequency is {\it greater} than the odd
one ({\it i.e.}, $m/n={\rm odd/even}$). We show that in the former
case the normal form of the coupled amplitude equations describing the
relevant resonant triad interaction does {\rm not} possess any
quadratic terms because of a symmetry associated with the subharmonic
waves.  This is in contrast to the case of odd/even forcing where
quadratic terms are present in the normal form.

In this paper, we demonstrate this distinction between even/odd and
odd/even forcing by considering a simple example of one-dimensional
waves that are parametrically excited by two frequencies. Rather than
consider the full hydrodynamic equations, we investigate this issue
using a simpler model derived from the free-surface Navier-Stokes
equation by Zhang and Vi\~{n}als~\cite{ref:zv97a}; their model applies
to a deep layer of low-viscosity fluid. We consider both
$2\omega/3\omega$ and $1\omega/2\omega$ forcing frequencies, over a
range of frequencies $\omega$, including a critical frequency for
which a one-dimensional spatial resonance occurs. Specifically, at the
critical frequency, the minima at the bicritical point are in ratio
$k_m/k_n=1/2$. We find that this spatial resonance leads to a
divergence of the Landau coefficient in the amplitude equation
describing bifurcation to subharmonic waves in the case of the $1/2$
forcing frequency ratio; this is consistent with the results of Zhang
and Vi\~{n}als~\cite{ref:zv97a}.  In contrast, the Landau coefficient
for waves, parametrically excited by two-frequency forcing in ratio
$2/3$, is unaffected by any parameter proximity to the bicritical
point and/or the spatial resonance point. While the wave vectors of
the critical modes ${\bf k}_1$ and $2{\bf k}_1$ are in resonance,
there is a mismatch in their frequencies: in this case the usual large
contribution to the Landau coefficient due to the spatial resonance is
absent.

The next section of the paper provides the necessary background to our
analysis.  It reviews the key theoretical ideas about the role of
resonant triad interactions in pattern formation problems, as well as
the linear theory for Faraday waves, parametrically excited by
two-frequency forcing. This background section closes with a
discussion of the normal form symmetries associated with the
subharmonic instability. Section~\ref{sec:model} sets up the example
of one--dimensional surface waves, indicating the precise analogy
between triad resonance in two dimensions and spatial resonance
within this simpler one--dimensional problem. This section also
presents the governing hydrodynamic equations that describe
parametrically excited surface waves in the limit of an infinite depth
fluid layer and weak viscosity. The last part of
Section~\ref{sec:model} sets up the weakly nonlinear analysis that we
use to compute the cubic Landau coefficient for harmonic and
subharmonic waves. Section~\ref{sec:results} presents bifurcation
results for two examples: waves parametrically excited by
two-frequency excitation in the ratio 2/3 and waves excited with a 1/2
excitation frequency ratio. Section~\ref{sec:conclude} summarizes our
results, and indicates some directions for future work.

\section{Background.}
\label{sec:background}

\subsection{Resonant triad interactions.}
\label{sec:resonant}

\begin{figure}
\centerline{
\epsfxsize=190pt
\epsffile{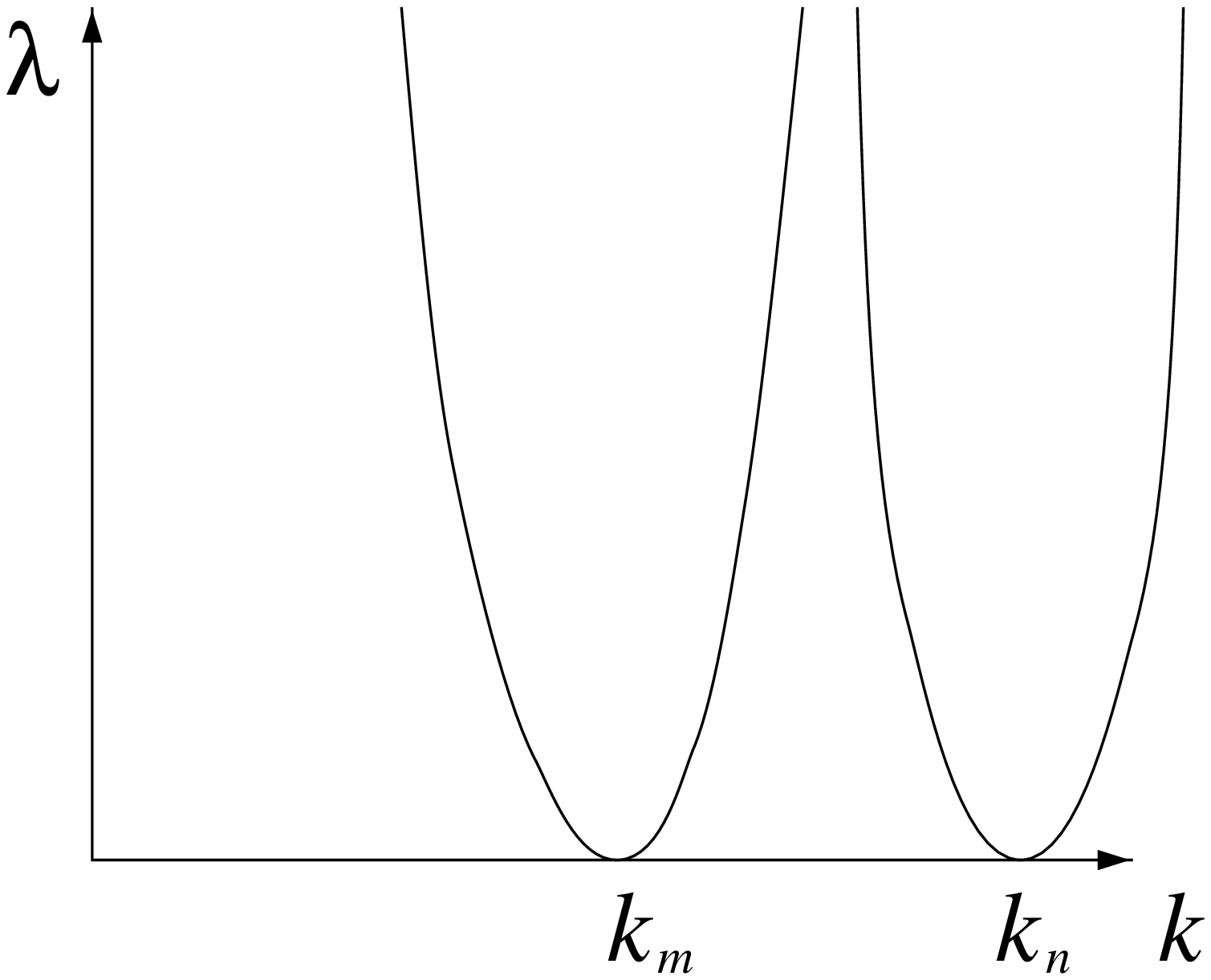}
\hskip 0.5truein
\epsfxsize=180pt
\epsffile{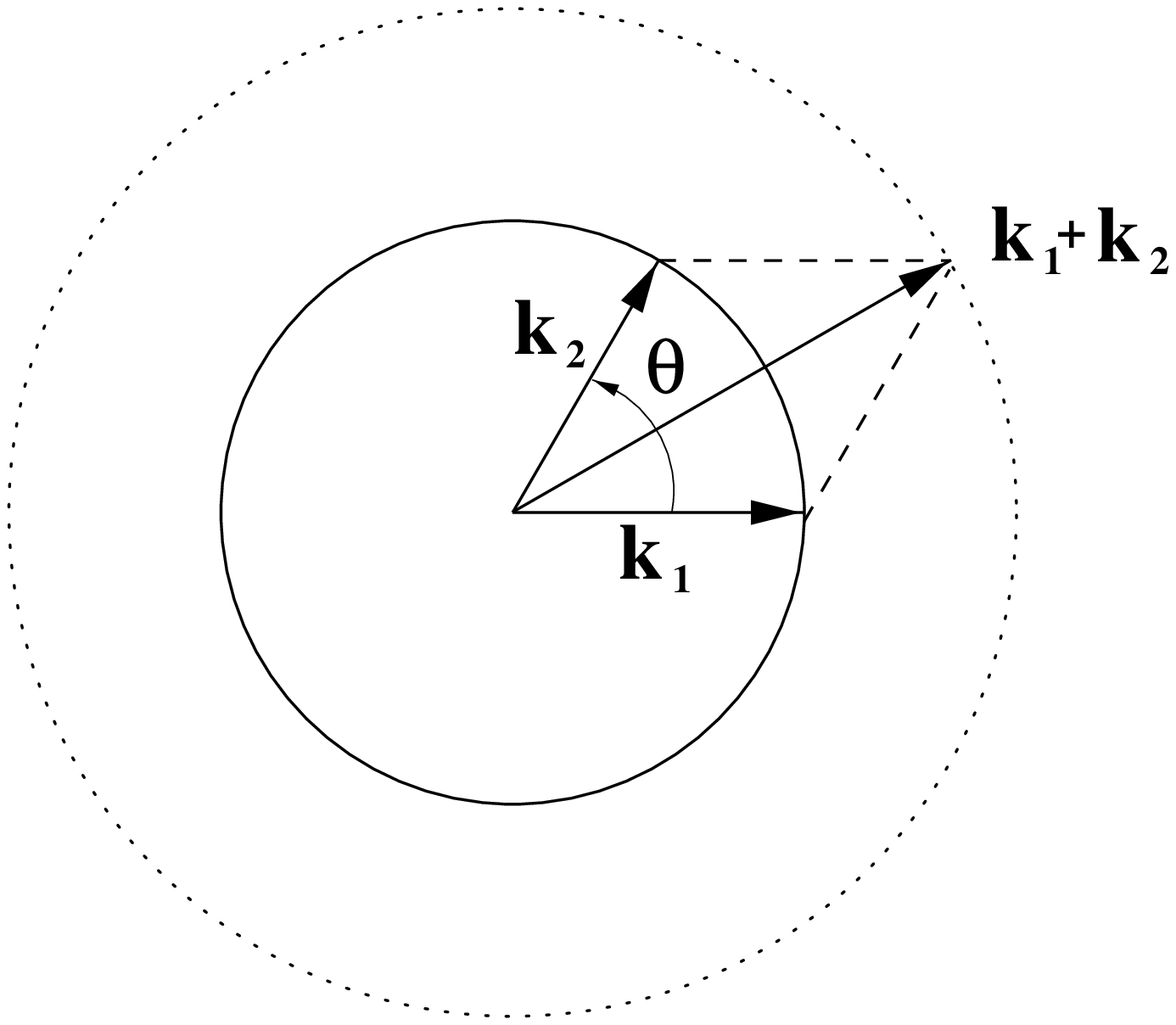}}
\caption{On the left is a plot of a neutral stability curve
$\lambda(k)$, showing minima at $k=k_m$ and $k=k_n$. On the
right is an associated resonant triad ${\bf k}_1$, ${\bf k}_2$ and
${\bf k}_1+{\bf k}_2$, where $|{\bf k}_1|=|{\bf k}_2|=k_m$ and
$|{\bf k}_1+{\bf k}_2|=k_n$. The angle $\theta$ between ${\bf k}_1$
and ${\bf k}_2$ is related to the ratio $k_n/k_m$ by (\protect\ref{angle}).}
\label{resonant-triad}
\end{figure}

One of the central ideas in pattern formation studies of
quasi-patterns~\cite{ef:94,ref:lp97,binks,newell}, and surface wave patterns
parametrically excited by two-frequency forcing~\cite{ref:zv97,ef:94} is that
of resonant triad interactions. In particular, there is interest in systems,
such as the Faraday system with two-frequency forcing, for which it is possible
to have a neutral stability curve with a double minimum. The resonant triad
interaction of interest is dictated by the locations of the minima, say $k_m$
and $k_n$, of the neutral curve. The resonant triad is made up of wave vectors
${\bf k}_1$, ${\bf k}_2$ and ${\bf k}_1+{\bf k}_2$, where $|{\bf k}_1|=|{\bf
k}_2|=k_m$ and $|{\bf k}_1+{\bf k}_2|=k_n$.  Thus the angle $\theta$ between ${\bf
k}_1$ and ${\bf k}_2$ satisfies
\begin{equation}
\label{angle}
\cos\Bigl({\theta\over 2}\Bigr)={k_n\over 2k_m}\ .
\end{equation} 
Here we have assumed that $0<k_m<k_n<2k_m$; see Figure~\ref{resonant-triad}.

To illustrate the possible influence of resonant triad interactions on
pattern formation, we first consider a steady state bifurcation
problem, where modes of wavenumber $k_m$ and $k_n$ lose stability
(almost) simultaneously with the increase of an external control
parameter. We consider a bifurcation problem on a six-dimensional
center manifold for linear modes
\begin{equation}
\label{eq:modes}
A(t)e^{i{\bf k}_1\cdot {\bf x}}+B(t)e^{i{\bf k}_2\cdot {\bf
x}}+C(t)e^{i({\bf k}_1+{\bf k}_2)\cdot {\bf x}}+c.c.,
\end{equation}
where $A,B,C\in{\bf C}$. Symmetry considerations determine that the unfolding
of the bifurcation problem takes the form, through cubic order in the
amplitudes,
\begin{eqnarray}
\label{ABC-steady}
\dot{A} &=& \lambda A+\alpha B^*C+(a|A|^2+b|B|^2+c|C|^2)A\nonumber\\
\dot{B} &=& \lambda B+\alpha A^*C+(a|B|^2+b|A|^2+c|C|^2)B\\
\dot{C} &=& \mu C+\beta AB+(d|A|^2+d|B|^2+e|C|^2)C\ ,\nonumber
\end{eqnarray}
where the asterisk indicates the complex conjugate, and the coefficients are
all real. If $\lambda=0$, $\mu\ne 0$, then one can further reduce the
bifurcation problem to one involving the critical modes $A$ and $B$, with $C$
constrained to the center manifold: $C=-{\beta\over \mu}AB+\cdots$. One
obtains, for $|\lambda|$ sufficiently small, the reduced (unfolded) bifurcation
problem
\begin{eqnarray}
\label{eq:amps}
\dot{A} &=& \lambda A+a|A|^2A+\Bigl(b-{\alpha\beta\over \mu}\Bigr)|B|^2A\nonumber\\
\dot{B} &=& \lambda B+a|B|^2B+\Bigl(b-{\alpha\beta\over \mu}\Bigr)|A|^2B\ .
\end{eqnarray}
The presence of the near critical mode $C$ in (\ref{eq:modes}) leads
to a large cross-coupling coefficient
$g(\theta)\equiv(b-{\alpha\beta\over\mu})$ in the amplitude equations
(\ref{eq:amps}) since $|\mu|\ll 1$.  A consequence of this is that
patterns that involve an equal amplitude superposition of modes $A$
and $B$ tend to be unstable at onset. For example, the stability of
steady rhombic patterns at angle $\theta$ ($A=B$ in (\ref{eq:amps})),
within the setting of the amplitude equations~(\ref{eq:amps}), is
determined by two eigenvalues whose signs are
$sgn(a+b-{\alpha\beta\over\mu})$ and $sgn(a-b+{\alpha\beta\over\mu})$. When
$|{\alpha\beta\over\mu}|\gg |a|+|b|$, the two eigenvalues take on opposite
signs and the pattern is necessarily unstable at onset. If the spatial
resonance occurs near $\theta=\pi/2$, for example, then square
planforms are unstable.  Quasi-patterns, and spatially-periodic
``superlattice'' patterns, such as described in
\cite{gollub,sp,dionne}, involving both of the modes $A$ and $B$ above
are similarly destabilized by a large cubic cross-coupling coefficient
$g(\theta)$ in the amplitude equations. This point is discussed for
various regular patterns, including quasi-patterns, in
\cite{ref:zv97}.

\subsection{Bicritical curves for the two-frequency Faraday problem.}
\label{subsec:linear}

We now examine, in greater detail, the double minimum of the neutral
curve associated with the Faraday problem with two-frequency
forcing. Besson, Edwards and Tuckerman \cite{laurette} computed the
linear stability of the flat free surface of a shallow fluid layer
subjected to a periodic vertical acceleration
\begin{equation}
\label{forcing}
g(t)=g_0+g_z[\cos(\chi)\cos(m\omega t)+\sin(\chi)\cos(n\omega
t+\phi)].
\end{equation} 
Here $g_0$ is the usual gravitational acceleration, $m$ and $n$ are
co-prime integers, and $g_z,\omega,\chi,\phi$ are additional external
control parameters. Note that the angle $\chi$ controls the relative
amplitudes of the $m\omega$--forcing and $n\omega$--forcing. Besson
{\it et al.} \cite{laurette} compared, with good agreement, their
theoretical predictions with their experimental results for various
fluids with different viscosities in the case that $m=4$ and $n=5$ in
(\ref{forcing}).  In this case, they found that the initial
instability of the fluid surface was either associated with a Floquet
multiplier (FM) $-1$ or $+1$, depending on whether the odd or even
frequency component in (\ref{forcing}) dominated. The transition
between subharmonic (FM$=-1$) and harmonic response (FM$=+1$) occurred
at a particular value of the angle $\chi$ in (\ref{forcing}), called
the bicritical point.

Here we do not consider the full hydrodynamic problem, but rather a
simplified model derived by Zhang and Vi\~{n}als \cite{ref:zv97a}, which
applies to a deep layer of weakly viscous fluid. The linearized
equation for the free surface mode $h_k(t)e^{ikx}$ is
\begin{equation}
\label{simple-lin}
h_k''+4\nu k^2 h_k'+\Bigl[g(t)k+{\Gamma k^3\over \rho}+4\nu^2 k^4\Bigr]h_k=0,
\end{equation}
where $\nu$ is the kinematic viscosity of the fluid, $\Gamma$ is the
surface tension, and $\rho$ is the fluid density. 
Note that in the absence of the time periodic parametric forcing ({\it i.e.}
$g_z=0$), one recovers a dispersion relation, where the damping is given by
$2\nu k^2$ and the water wave frequency $\omega_0$ satisfies the usual
gravity-capillary wave dispersion relation $\omega_0^2=g_0k+\Gamma
k^3/\rho$.

\begin{figure}
\vskip 5.5truein
\hskip -1truein
\centerline{
\epsfxsize=225pt
\epsffile{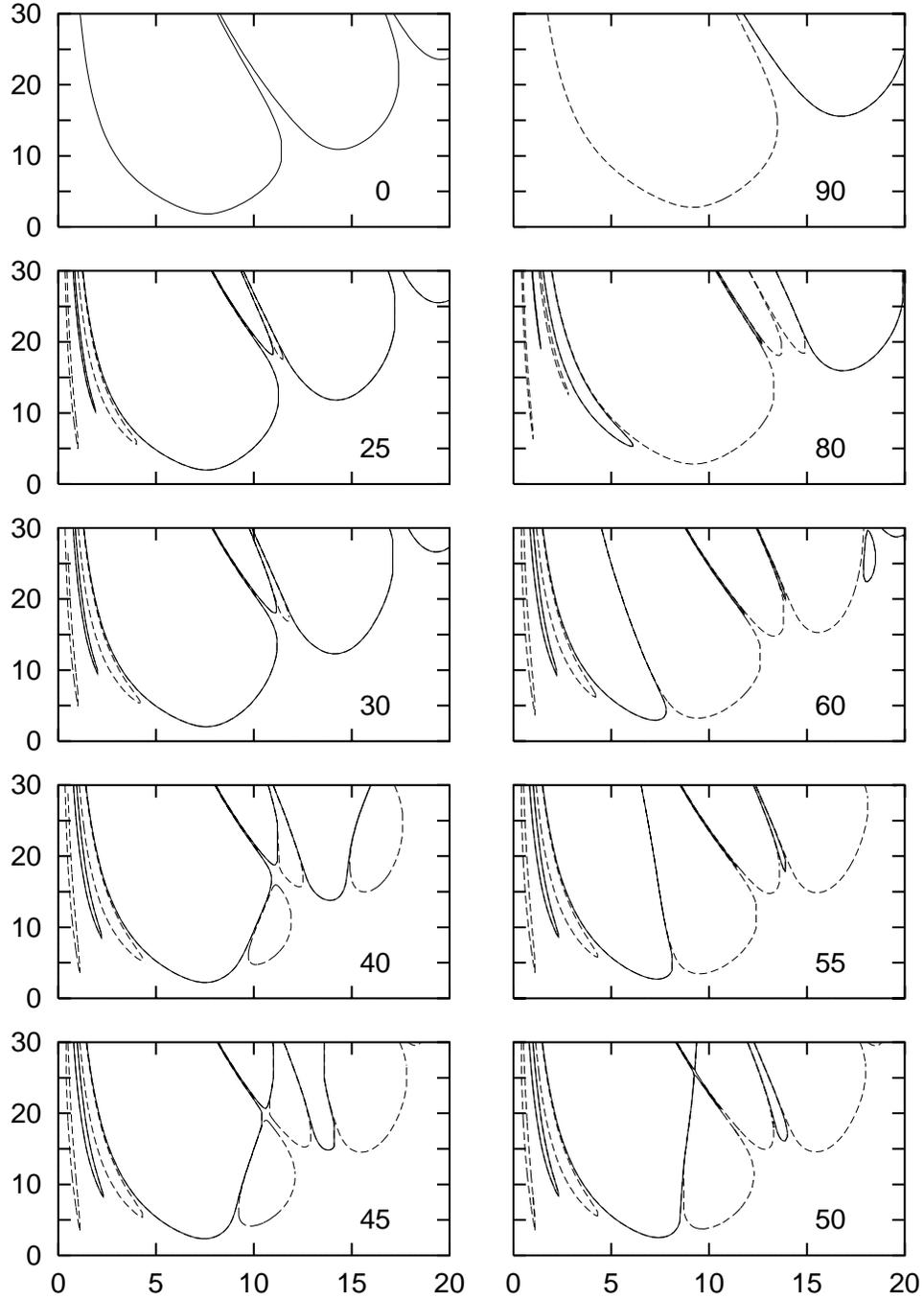}}
\vskip -0.7truein
\caption{Neutral stability curves computed
from~(\protect\ref{simple-lin}).  The vertical axis is $g_z/g_0$ and
the horizontal axis is $k$; the angle $\chi$
in~(\protect\ref{forcing}) is indicated in the lower right corner of
each plot.  Floquet multipliers of $+1$ ($-1$) are indicated by solid
(dashed) lines. The first instability encountered with increased
forcing $g_z$ is to harmonic waves if $\chi<\chi_c\approx 63^\circ$,
and to subharmonic waves for $\chi>\chi_c$. The other parameters of
the forcing function are $m=4$, $n=5$, $\phi=0$ and $\omega/2\pi=11\
Hz$; the fluid parameter are $\Gamma=20.6\ dyn/cm$, $\nu=0.209\
cm^2/s$ and $\rho=0.95\ g/cm^3$.}
\label{neutral-curve}
\end{figure}

In Figure~\ref{neutral-curve} we present an example of the neutral stability
curves $g_z(k)$ for the parameters used by Besson, {\it et al.}
\cite{laurette}, and for various values of the parameter $\chi$. These curves
were obtained from (\ref{simple-lin}) using the same method that
Besson, {\it et al.} \cite{laurette} used to obtain the neutral curves
for the full hydrodynamic problem. The bicritical point in this
example occurs at $\chi=63.11^\circ$, for $g_z=3.13\ g_0$
($g_0=980.665\ cm/s^2$), $k_+=7.14\ cm^{-1}$ and $k_-=9.35\ cm^{-1}$,
where $k_\pm$ are the values of $k$ associated with FMs $\pm 1$. The
locations of the minima can be roughly approximated by considering the
simple water wave dispersion relation $\omega_0^2=g_0k+\Gamma
k^3/\rho$, and assuming subharmonic response to the forcing
frequencies at $m\omega$ and $n\omega$, {\it i.e.}, for $m=4$ and
$n=5$:
\begin{eqnarray}
\label{simple}
\Bigl({4\omega\over 2}\Bigr)^2&\approx &g_0k_++{\Gamma
k_+^3\over \rho}\nonumber\\
\Bigl({5\omega\over 2}\Bigr)^2&\approx &g_0k_-+{\Gamma
k_-^3\over\rho}
\end{eqnarray}
For the example of Figure~\ref{neutral-curve} this approximation yields
$k_+\approx 8.0$ and $k_-\approx 9.8$.  We note that the neutral stability
curves obtained from the simple linear problem (\ref{simple-lin}) are
qualitatively very similar to those obtained by Besson, {\it et
al.}~\cite{laurette} from the full hydrodynamic problem, {\it cf.}  our
Figure~\ref{neutral-curve} with Figure 2 in \cite{laurette}.

\subsection{Normal form symmetries.}

We now re-examine the role of resonant triad interactions for the mode
interaction problem pertinent to the two-frequency Faraday experiment.
In particular, we will investigate the additional effect of normal
form symmetries on the pattern formation problem. We wish to contrast
the situations where the forcing frequencies $m\omega$ and
$n\omega$ in (\ref{forcing}), $m<n$, have $m$ even, $n$ odd with $m$
odd, $n$ even. We are specifically interested in the case where the
angle $\chi$ in (\ref{forcing}) is close to the bicritical point
$\chi_c$. 

We analyze the resonant triad interaction in terms of a stroboscopic
map since we are interested in a periodically-forced system.  We
denote the surface height at time $t=pT$, $p$ an integer,
$T=2\pi/\omega$, by $h_p({\bf x})$. Let the resonant triad be
\begin{equation}
h_p=A_pe^{i{\bf k}_1\cdot{\bf x}}+B_pe^{i{\bf k}_2\cdot{\bf
x}}+C_pe^{i({\bf k}_1+{\bf k}_2)\cdot{\bf x}}+c.c.+\cdots,
\end{equation}
where $|{\bf k}_1|=|{\bf k}_2|=k_m$ and $|{\bf k}_1+{\bf k}_2|=k_n$,
$0<k_m<k_n<2k_m$. Here the $m,n$ subscripts indicate that the critical
wavenumbers can be roughly associated with the $m\omega$ and $n\omega$
forcing, respectively, as in (\ref{simple}) above.  The general form
of the cubic--order amplitude equations, consistent with translation
symmetry and reflection symmetry, is ({\it cf.} equation
\ref{ABC-steady})
\begin{eqnarray}
\label{ABC-mode-interaction}
{A}_{p+1} &=& \lambda A_p+\alpha
B_p^*C_p+(a|A_p|^2+b|B_p|^2+c|C_p|^2)A_p
\nonumber\\
{B}_{p+1} &=& \lambda B_p+\alpha
A_p^*C_p+(a|B_p|^2+b|A_p|^2+c|C_p|^2)B_p\\
{C}_{p+1} &=& \mu C_p+\beta
A_pB_p+(d|A_p|^2+d|B_p|^2+e|C_p|^2)C_p\ .\nonumber
\end{eqnarray}
In the case that $m$ is even and $n$ is odd, the Floquet multipliers
for the linearized problem are $\lambda=+1$ and $\mu=-1$ when
$\chi=\chi_c$, whereas if $m$ is odd and $n$ is even, then
$\lambda=-1$, $\mu=+1$. 

It may be possible to further simplify the bifurcation problem
(\ref{ABC-mode-interaction}) by a normal form transformation. In particular,
there exists a near-identity nonlinear transformation such that all nonlinear
terms in (\ref{ABC-mode-interaction}), which do not commute with the matrix $L$
of the linearized problem, can be removed (see, for example, Crawford
\cite{crawford}). Here
\begin{equation}
L=\pmatrix{\lambda & 0 &0\cr 0 & \lambda &0 \cr 0 & 0 & \mu},
\end{equation}
where $|\lambda|=|\mu|=1$. 

In the case that $\lambda= -1$, $\mu=+1$, {\it i.e.}, $m$ odd and $n$ even in
the forcing (\ref{forcing}), the bifurcation problem
(\ref{ABC-mode-interaction}) is already in normal form since the normal form
symmetry is equivalent to a translation by ${\bf d}$, where ${\bf k}_1\cdot
{\bf d}={\bf k}_2\cdot {\bf d}=\pi$.  The presence of the quadratic terms in
this normal form means that the resonant triad interactions will have a strong
influence on the pattern formation problem, as described for the simple steady
state bifurcation example in Section~\ref{sec:resonant}. 

In contrast, in the case that $\lambda=+1$, $\mu=-1$, {\it i.e.}, $m$
even and $n$ odd, the normal form transformation allows the quadratic
terms in the bifurcation problem to be removed. The normal form of the
bifurcation problem, through cubic order, is
\begin{eqnarray}
\label{ABC-pm-nf}
{A}_{p+1} &=& A_p+(a|A_p|^2+b|B_p|^2+c|C_p|^2)A_p
\nonumber\\
{B}_{p+1} &=& B_p+(a|B_p|^2+b|A_p|^2+c|C_p|^2)B_p\\
{C}_{p+1} &=& - C_p+(d|A_p|^2+d|B_p|^2+e|C_p|^2)C_p\ .\nonumber
\end{eqnarray}
In this case, the bifurcation to harmonic waves is investigated in the
invariant subspace $C=0$, and there is no divergence of the cross-coupling term
due to the resonant triad interaction.  The usual contribution of the resonant
triad interaction to the pattern formation problem is suppressed in this
case. In the next section we test these ideas by performing explicit
computations of bifurcation coefficients from the simplified hydrodynamic
model, due to Zhang and Vi\~{n}als \cite{ref:zv97a}, of the two-frequency
Faraday experiment.
 
\section{Two-frequency forcing of one-dimensional surface waves.}
\label{sec:model}

\subsection{Resonant interactions in a one-dimensional problem.}
\label{sec-res-one}

It is possible to investigate the effects of strong spatial resonance
on a pattern formation problem in one spatial dimension. This is done
by considering the situation where the minima $k_m$ and $k_n$ of the
neutral curve satisfy $k_m/k_n=1/2$.  In this case, the stroboscopic
map associated with the critical modes $Z_pe^{ikx}+W_pe^{2ikx}+c.c.$
is, through cubic order,
\begin{eqnarray}
\label{zw-1d}
{Z}_{p+1} &=& \lambda Z_p+\alpha W_p Z_p^{*}+ (a|Z_p|^2+b|W_p|^2)Z_p
\nonumber\\
{W}_{p+1} &=& \mu W_p+\beta Z_p^2+(c|W_p|^2+d|Z_p|^2)W_p\ .
\end{eqnarray}
In form, this is identical to (\ref{ABC-mode-interaction}) restricted
to the subspace $A=B=Z$, $C=W$. 
 
If $\lambda=-1$ and $\mu=+1$, then (\ref{zw-1d}) is already in normal
form. The unfolding is
\begin{eqnarray}
\label{zw-1d-nf2}
{Z}_{p+1} &=& -(1+\epsilon) Z_p+\alpha W_p Z_p^{*}+ (a|Z_p|^2+b|W_p|^2)Z_p
\nonumber\\
{W}_{p+1} &=& (1+\delta) W_p+\beta Z_p^2+(c|W_p|^2+d|Z_p|^2)W_p\ .
\end{eqnarray}
Subharmonic waves bifurcate at $\epsilon=0$. For $\delta\ne 0$, the
center manifold is given by $W=-\beta Z^2/\delta+\cdots$, and the branching
equation is then given by
\begin{equation}
0=-\epsilon+(a-\alpha\beta/\delta)|Z|^2\ .
\end{equation}
We see that the presence of the near critical spatially-resonant
harmonic mode $W$ forces the subharmonic wave to bifurcate with very
small amplitude, since the cubic coefficient is large when $\delta$ is
small. This effect is the one-dimensional analogue of the cross-coupling term
$g(\theta)$ diverging near a spatial resonance.  This is the situation
that can occur when $m$ is odd and $n$ is even ($n>m$) in
(\ref{forcing}), {\it e.g.} $m=1,n=2$.

If, on the other hand, $\lambda=+1$, $\mu=-1$, then the quadratic
terms in (\ref{zw-1d}) can be removed, and the usual effect of the
spatial resonance on the pattern formation problem is suppressed.
Specifically, the (unfolded) normal form of the bifurcation problem is
\begin{eqnarray}
\label{zw-1d-nf}
{Z}_{p+1} &=& (1+\epsilon) Z_p+ (a|Z_p|^2+b|W_p|^2)Z_p
\nonumber\\
{W}_{p+1} &=& -(1+\delta) W_p+(c|W_p|^2+d|Z_p|^2)W_p\ .
\end{eqnarray}
Harmonic waves, which bifurcate from the trivial solution at
$\epsilon=0$, are investigated in the invariant subspace $W=0$; they
satisfy the branching equation
\begin{equation}
0=\epsilon+a|Z|^2\ .
\end{equation}
Subharmonic waves bifurcate from the trivial solution at $\delta=0$
and can be investigated in the invariant subspace $Z=0$; they
satisfy the branching equation
\begin{equation}
0=-\delta+c|W|^2\ .  
\end{equation}
Neither of these branches is affected by the spatial resonance.  This
is the situation we expect when $m$ is even and $n$ is odd in
(\ref{forcing}), {\it e.g.} $m=2$, $n=3$. 

\begin{figure}
\vskip 1.9truein
\hskip 0.1truein
\centerline{
\epsfxsize=215pt
\epsffile{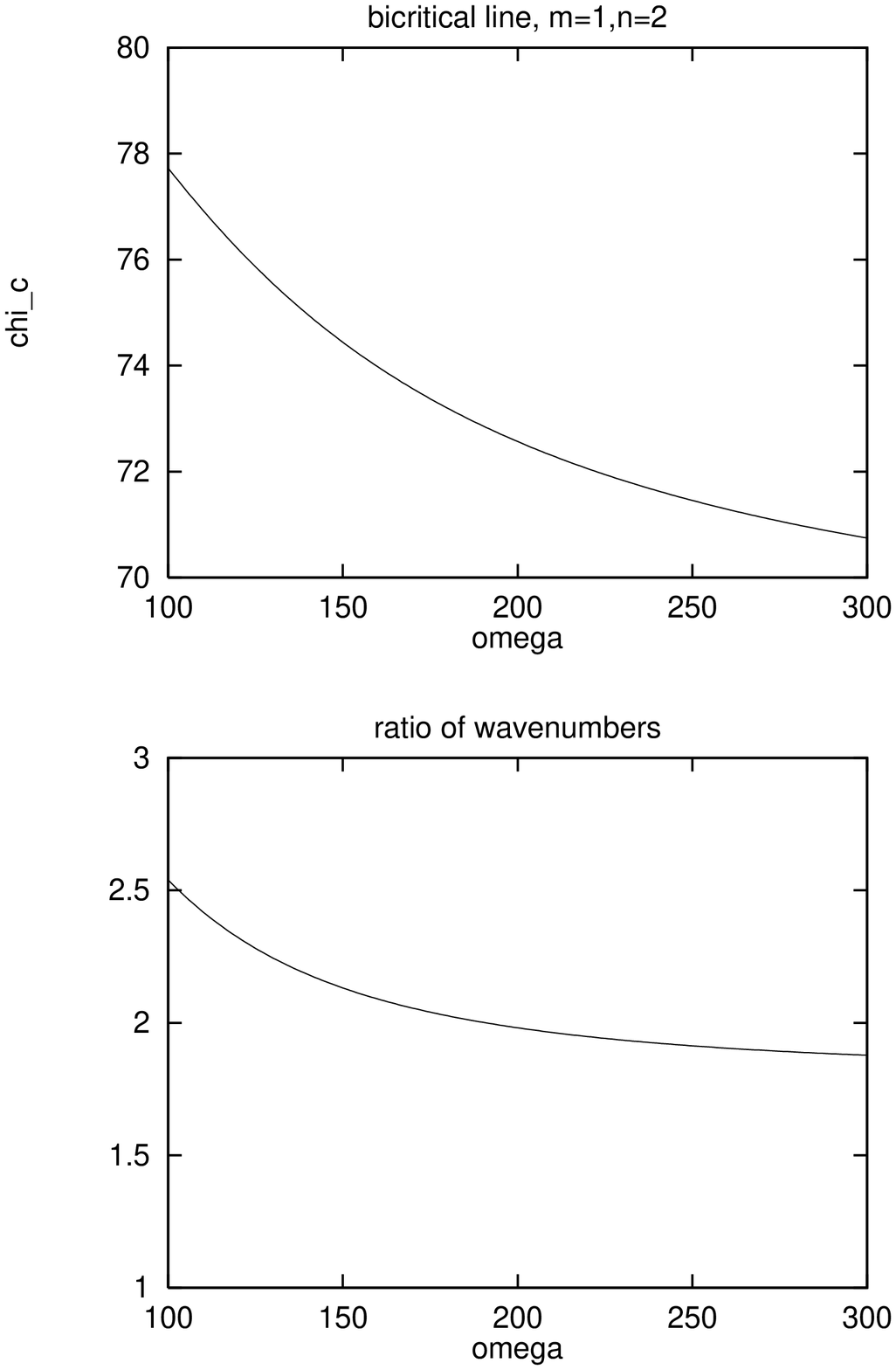}
\hskip -0.2truein
\epsfxsize=215pt
\epsffile{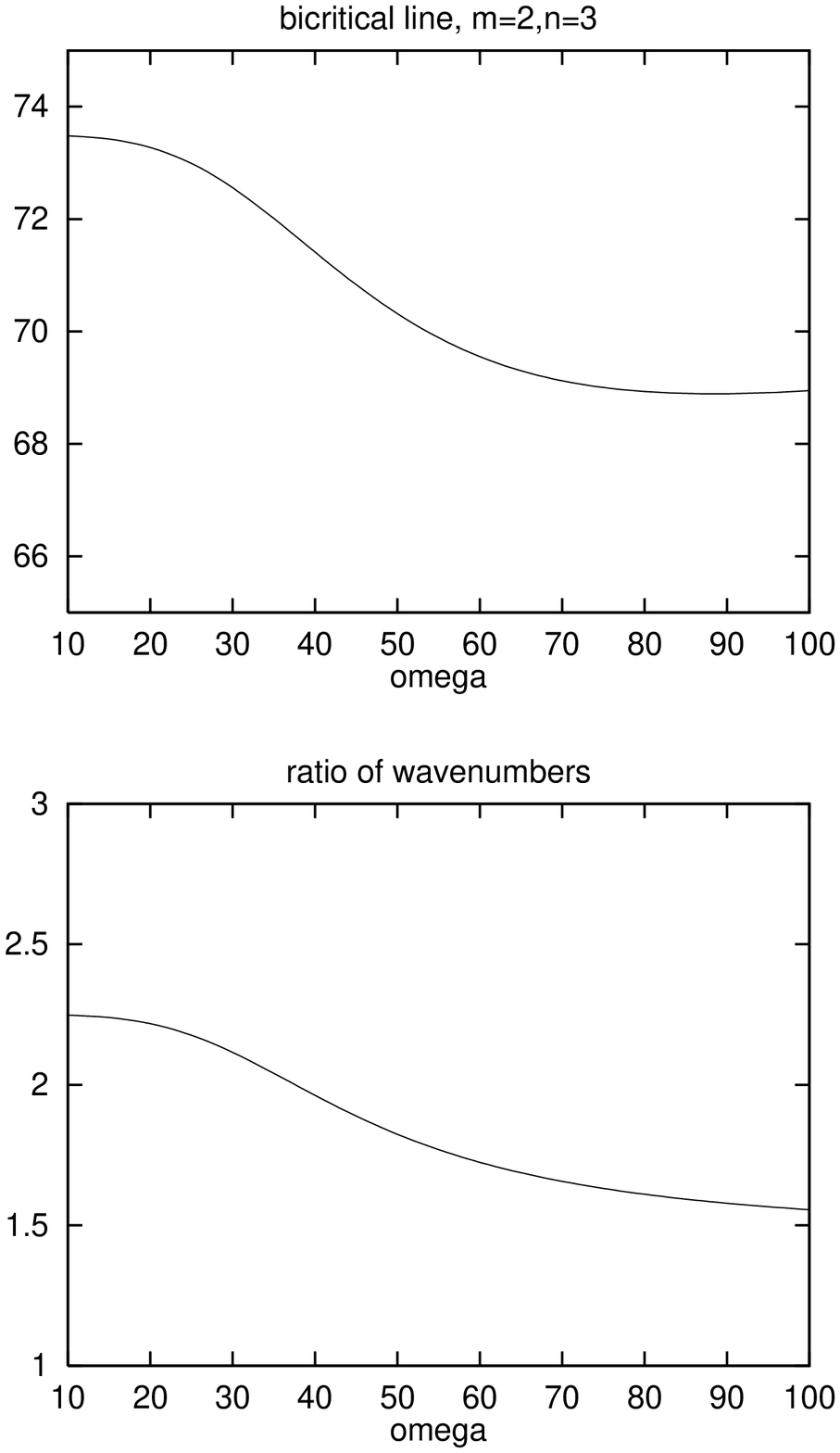}}
\vskip -0.45truein
\caption{Bicritical point $\chi_c$ plotted as a function of the frequency
$\omega$ in~(\protect\ref{forcing}), where $\phi=0$ and $\Gamma=20.6\
dyn/cm$, $\nu=0.209\ cm^2/s$ and $\rho=0.95\ g/cm^3$
in~(\protect\ref{simple-lin}).  Plotted below each bicritical line is
the ratio of critical wavenumbers $k_n/k_m$ for
$\chi=\chi_c(\omega)$. On the left $m=1,n=2$, and on the right $m=2,
n=3$.}
\label{bi-critical}
\end{figure}

In Figure~\ref{bi-critical}, we plot the bicritical point $\chi_c$ as
a function of $\omega$ for both $(m,n)=(2,3)$ and $(m,n)=(1,2)$. These
curves apply to the linear equation (\ref{simple-lin}), with
$\Gamma=20.6\ dyn/cm$, $\nu=0.209\ cm^2/s$ and $\rho=0.95\ g/cm^3$. In
this figure we show also the ratio of wavenumbers $k_n/k_m$ at the
bicritical point. Note that in each case the spatial resonance
$k_n/k_m=2$ is achieved for a particular value of the forcing
frequency $\omega=\omega^{res}$.

\subsection{The Zhang-Vi\~{n}als hydrodynamic model.}

Zhang and Vi\~{n}als \cite{ref:zv97} derive the following model for the
surface deviation $h({\bf x},t)$ and surface velocity potential
$\widehat\Phi({\bf x},t)$, where ${\bf x}\in {\bf R}^2$:
\begin{eqnarray}
\label{eq:basic}
(\partial_t-\widehat\gamma\nabla^2)h-\widehat{\cal D}\widehat\Phi
&=&\widetilde{\cal F}(h,\widehat\Phi)\\
(\partial_t-\widehat\gamma\nabla^2)\widehat\Phi-\widehat{\cal A}h&=&
\widetilde{\cal G}(h,\widehat\Phi)\
.\nonumber
\end{eqnarray}
Here the linear operator $\widehat{\cal A}$ is
\begin{equation}
\widehat{\cal A}\equiv \widehat\Gamma_0\nabla^2-\widehat G_0-4\widehat
f[\sin(2t)+\kappa\sin(2pt+\phi)],
\nonumber
\end{equation}
where $p=n/m$. The linear, nonlocal operator $\widehat{\cal D}$
multiplies each Fourier component of the field by its wavenumber
modulus. $\widetilde{\cal F}$ and $\widetilde{\cal G}$ are 
nonlinear operators to be
defined shortly, together with the parameters $\widehat\gamma$, 
$\widehat\Gamma_0$,
$\widehat G_0$, $\widehat f$ and $\kappa$. 

In deriving the above, Zhang and Vi\~{n}als assumed that the experimental
forcing takes the form $g_zr(\sin(2m\omega_0 t)+\kappa\sin(2n\omega_0
t+\phi))$, where $\kappa\equiv(1-r)/r$; they have scaled time by the
frequency $m\omega_0$. Here we choose a slightly different scaling of
time; we assume that the parametric forcing takes the (dimensionless) form
$f[\cos(\chi)\cos(m\tau)+\sin(\chi)\cos(n\tau+\phi)]$. We let
$t=m\tau/2$ and $\widehat\Phi={2\over m}\Phi$ in (\ref{eq:basic}) to
obtain
\begin{eqnarray}
\label{eq:basic2}
(\partial_\tau-\gamma\nabla^2)h-\widehat{\cal D}\Phi
&=&{\cal F}(h,\Phi)\\
(\partial_\tau-\gamma\nabla^2)\Phi-{\cal A}h
&=&{\cal G}(h,\Phi)\ ,\nonumber
\end{eqnarray}
where 
\begin{equation}
\gamma\equiv {2\nu k_0^2\over \omega}.
\end{equation}
The wavenumber $k_0$ satisfies the equation
\begin{equation}
g_0k_0+{\Gamma k_0^3\over \rho}={m^2\omega^2\over 4},
\end{equation}
where $\omega=2\omega_0$ is the dimensioned forcing frequency.  The
linear operator ${\cal A}$ is now
\begin{equation}
\label{a-operator}
{\cal A}\equiv \Gamma_0\nabla^2- G_0-
f(\cos(\chi)\cos(m\tau)+\sin(\chi)\cos(n\tau+\phi)),
\end{equation}
where 
\begin{equation}
\Gamma_0\equiv {\Gamma k_0^3\over \rho\omega^2}, \quad
G_0\equiv{gk_0\over \omega^2},\quad
f\equiv {g_zk_0\over \omega^2}\ .\nonumber
\end{equation}
The nonlinear operators are
\begin{eqnarray}
{\cal F}(h,\Phi)&\equiv& -\nabla\cdot(h\nabla
\Phi)+{1\over 2}\nabla^2(h^2\widehat{\cal D}\Phi)
-\widehat{\cal D}(h\widehat{\cal D} \Phi)+\widehat{\cal
D}[h\widehat{\cal D}(h\widehat{\cal D}\Phi)+{1\over
2}h^2\nabla^2{\Phi}]\\
{\cal G}(h,\Phi)&\equiv  & {1\over 2}(\widehat
D\Phi)^2-{1\over 2}(\nabla \Phi)^2-(\widehat{\cal D}
\Phi)[h\nabla^2\Phi+\widehat{\cal D}(h\widehat{\cal
D}\Phi)]-{1\over 2}\Gamma_0\nabla\cdot(\nabla h(\nabla h)^2)).
\nonumber
\end{eqnarray}

Finally, we put the governing equations in a convenient form for the
perturbation analysis that follows. In order to recover the linearized
equation (\ref{simple-lin}) for $h$, we apply the operator
$(\partial_\tau-\gamma\nabla^2)$ to the first of equations (\ref{eq:basic2}),
and then use the second equation to express
$(\partial_\tau-\gamma\nabla^2)\Phi$ as ${\cal A}h+{\cal G}(h,\Phi)$. We
supplement this equation with the first of equations (\ref{eq:basic2}),
re-written as an equation for the field $\Phi$, to obtain the full system of
equations:
\begin{eqnarray}
\label{eq:basic3}
{\cal L}h&=&(\partial_\tau-\gamma\nabla^2){\cal F}(h,\Phi)+
\widehat{\cal D}{\cal
G}(h,\Phi)\\
\widehat{\cal D}\Phi&=&(\partial_\tau-\gamma\nabla^2)h-{\cal
F}(h,\Phi)\ .\nonumber
\end{eqnarray}
Here the linear operator is
\begin{equation}
\label{eq:linear}
{\cal L}\equiv \partial_{\tau\tau}-2\gamma\nabla^2\partial_\tau
+(\gamma^2\nabla^4-\widehat {\cal D}{\cal A})\ ,  
\end{equation}
where the operator ${\cal A}$ is given by~(\ref{a-operator}).

\subsection{Weakly nonlinear analysis: derivation of the bifurcation problem.}

We restrict our analysis to spatially one-dimensional solutions of the
variable $x\in{\bf R}$, and use a two-timing perturbation method to
determine weakly nonlinear solutions of the system (\ref{eq:basic3}) in the
vicinity of the bifurcation to (sub)harmonic waves, {\it i.e.} near a
primary instability associated with a Floquet multiplier $+1$ ($-1$).
To do this we introduce a small parameter $\epsilon$, such that
\begin{eqnarray}
h(x,\tau)&=&\epsilon h_1(x,\tau,T)+\epsilon^2 h_2(x,\tau,T)+
\epsilon^3 h_3(x,\tau,T)+\cdots\\
\Phi(x,\tau)&=&\epsilon \Phi_1(x,\tau,T)+\epsilon^2
\Phi_2(x,\tau,T)+\epsilon^3
\Phi_3(x,\tau,T)+\cdots\ ,
\nonumber
\end{eqnarray}
where
\begin{equation}
T=\epsilon^2\tau\ , \quad f=f_0+\epsilon^2 f_2\ .
\end{equation}
Here $f_0$ is the value of the forcing at the bifurcation point. We
seek spatially-periodic solutions in the following separable
Floquet-Fourier form:
\begin{eqnarray}
h_1=z_1(T)p_1(\tau)e^{ik_cx}+c.c.&\quad&
h_2=z_1^2(T)p_2(\tau)e^{i2k_cx}+c.c.\\
\Phi_1=z_1(T)q_1(\tau)e^{ik_cx}+c.c.&\quad&
\Phi_2=z_1^2(T)q_2(\tau)e^{i2k_cx}+c.c.\nonumber
\end{eqnarray}
Here $p_l, q_l$ ($l=1,2$) are $2\pi$-periodic functions of the fast
time $\tau$ in the case of harmonic waves (FM=+1); in the case of
subharmonic waves they are $4\pi$-periodic in $\tau$. The wavenumber
$k_c$ is associated with the first unstable mode.

At leading order in $\epsilon$, we obtain
\begin{eqnarray}
\label{eq:orderepsilon}
{\cal L}_0h_1&=&0\\
\widehat{\cal D}\Phi_1&=&(\partial_\tau-\gamma\partial_{xx})h_1\nonumber
\end{eqnarray}
Here the linear operator ${\cal L}_0$ is ${\cal L}$ with $f=f_0$, where
${\cal L}$ is defined by (\ref{eq:linear}). The equation
${\cal L}_0h_1=0$ determines the bifurcation point $f_0$ by the solvability
condition that it have a periodic solution; it is the equation we solved in
Section~\ref{subsec:linear} to determine the neutral stability curves. This
equation also determines the critical wavenumber $k_c$ by the requirement that
$f_0$ be the smallest value of $f$ that admits a periodic solution. Given
$f_0$ and $k_c$, we determine the solution of ${\cal L}_0h_1=0$ involving the
periodic function $p_1(\tau)$, which we may assume is real.
We express $p_1$ in terms of its (truncated)
Fourier series
\begin{equation}
\label{eq:p1}
p_1(\tau)=\sum_{j=0}^N a_je^{i(j+\mu)\tau}+c.c.,
\end{equation}
where $N$ is chosen large enough so that the solution is well converged. Here
$\mu=0$ for the case of harmonic waves and $\mu=1/2$ for subharmonic waves.
We readily solve the second of equations (\ref{eq:orderepsilon}) for $\Phi_1$,
yielding
\begin{equation}
q_1(\tau)={1\over k_c}(\partial_\tau+\gamma k_c^2)p_1\ .
\end{equation}

At order $\epsilon^2$, we obtain
\begin{eqnarray}
\label{eq:order2}
{\cal
L}_0h_2&=&-(\partial_\tau-\gamma\partial_{xx})[\partial_x(h_1\partial_x
\Phi_1)+\widehat{\cal D}(h_1\widehat{\cal D}\Phi_1)]
+{1\over 2}\widehat{\cal D}[(\widehat{\cal D}\Phi_1)^2
-(\partial_x\Phi_1)^2]\\
&=&2k_c^3z_1^2q_1^2e^{2ik_cx}+c.c.,\nonumber
\end{eqnarray}
where $q_1^2$ is a real $2\pi$--periodic function. We
determine a solution $h_2$ involving the real periodic function 
$p_2(\tau)$, given by
its (truncated) Fourier series representation
\begin{equation}
p_2=\sum_{j=0}^N b_je^{ij\tau}+c.c.
\end{equation}
We also find, at this order, that $\Phi_2$ satisfies
\begin{equation}
\widehat{\cal D}\Phi_2=(\partial_\tau-\gamma\partial_{xx})h_2
\end{equation}
since $\partial_x(h_1\partial_x \Phi_1)+\widehat{\cal D}(h_1\widehat{\cal D}
\Phi_1)=0$; thus 
\begin{equation}
q_2(\tau)={1\over 2k_c}(\partial_\tau+4\gamma k_c^2)p_2\ .
\end{equation}

Finally, we consider the equation at order $\epsilon^3$:
\begin{eqnarray}
{\cal
L}_0h_3&=&-2\partial_\tau\partial_Th_1+2\gamma\partial_{xx}\partial_Th_1
-f_2[\cos(\chi)\cos(m\tau)+\sin(\chi)\cos(n\tau+\phi)]\widehat{\cal
D}h_1\nonumber\\
&&+(\partial_\tau-\gamma\partial_{xx})\Bigl[-\partial_x(h_1\partial_x\Phi_2
+h_2\partial_x\Phi_1)-\widehat{\cal D}(h_1\widehat{\cal
D}\Phi_2+h_2\widehat{\cal D}\Phi_1)\\
&&+{1\over 2}\partial_{xx}(h_1^2\widehat{\cal D}\Phi_1)+\widehat{\cal
D}[h_1\widehat{\cal D}(h_1\widehat{\cal D}\Phi_1)+{1\over
2}h_1^2\partial_{xx}\Phi_1]\Bigr]+\widehat{\cal D}\Bigl[(\widehat{\cal
D}\Phi_1)(\widehat{\cal D}\Phi_2)\nonumber\\
&&-(\partial_x\Phi_1)(\partial_x\Phi_2)-\widehat{\cal
D}\Phi_1[h_1\partial_{xx}\Phi_1+\widehat{\cal D}(h_1\widehat{\cal
D}\Phi_1)]
-{\Gamma_0\over
2}\partial_x(\partial_xh_1)^3\Bigr]\nonumber\ .
\end{eqnarray}
In order to ensure that a $2\pi$-periodic solution exists, we must
apply a solvability condition to this equation, written
compactly
as ${\cal L}_0h_3=H(x,\tau,T)$. Specifically, we require
\begin{equation}
\langle \widetilde h_1,{\cal L}_0h_3\rangle=\langle 
\widetilde h_1,H(x,\tau,T)\rangle=0,
\end{equation}
where the inner product is
\begin{equation}
\langle f,g\rangle\equiv{k_c\over 8\pi^2}\int_0^{4\pi}
d\tau\int_0^{2\pi/k_c}  f^*(x,\tau)g(x,\tau) dx;
\end{equation}
$\widetilde h_1\equiv \widetilde p_1(\tau)e^{ik_c x}$ 
is a periodic solution to the adjoint linear problem
${\cal L}_0^\dagger\widetilde h_1=0$. Here
\begin{equation}
{\cal L}_0^\dagger\equiv
(\partial_{\tau}+\gamma\partial_{xx})^2-\widehat{\cal D}{\cal A},
\end{equation}
where $f=f_0$ in ${\cal A}$.  

The solvability condition leads to the amplitude equation
\begin{equation}
\label{landau}
\delta{dz_1\over dT}=\alpha f_2 z_1+(A+B)|z_1|^2z_1,
\end{equation}
where
\begin{eqnarray}
\label{integrals}
\delta&=&{1\over 2\pi}\int_0^{4\pi}(p_1'+i\mu p_1+\gamma
k_c^2p_1) \widetilde p_1\ d\tau\nonumber\\
\alpha&=&{-k_c\over 4\pi}\int_0^{4\pi}
[\cos(\chi)\cos(m\tau)+\sin(\chi)
\cos(n\tau+\phi)]p_1\widetilde p_1\  d\tau\\
A&=&-{k_c^2\over 2\pi}\int_0^{4\pi} ((q_1p_2)'+i\mu q_1p_2
+\gamma k_c^2q_1p_2)\widetilde p_1 \ d\tau\nonumber\\
B&=&{k_c^3\over 4\pi}\int_0^{4\pi} (-(p_1^2q_1)'-i\mu p_1^2q_1
-\gamma k_c^2 p_1^2q_1+k_c q_1^2p_1+{3\over 2}k_c^2\Gamma_0|p_1|^2p_1)
\widetilde p_1\ d\tau.\nonumber
\end{eqnarray}
In the above, differentiation with respect to $\tau$ is denoted by a prime.
We also took the periodic solution $\widetilde p_1$ to be real. 

In the above derivation of the bifurcation problem~(\ref{landau}), we
have separated the two contributions $A$ and $B$ to the cubic
coefficient $A+B$.  The contribution $A$ comes from the quadratic
nonlinear terms in the original hydrodynamic model; it depends on the
spatially resonant modes $e^{i2k_cx}$ as is evident from the $p_2$
terms in the integral expression for $A$ given
in~(\ref{integrals}). The contribution $B$ comes from cubic
nonlinearities and therefore depends only on the mode $e^{ik_cx}$. 
We calculate all of the coefficients $\delta, \alpha, A$ and $B$
numerically as follows. 

At leading order in $\epsilon$, equation (\ref{eq:orderepsilon}) for
$h_1$ reduces to the linear problem (\ref{simple-lin}). The solution for
$h_1$ is found by representing $p_1(\tau)$ as the truncated Fourier
series given in equation (\ref{eq:p1}).  As discussed in
\cite{laurette}, the problem then reduces to a generalised eigenvalue
problem of the form
\begin{equation}
{\bf A x} = g_z {\bf B x},\nonumber
\end{equation}
where ${\bf x}$ is a vector of the Fourier coefficients $a_j$ in
(\ref{eq:p1}). The EISPACK routine {\tt rgg} is used to find the
eigenvalues and corresponding eigenvectors, the eigenvalues giving the
linear stability curves shown in Figure~\ref{neutral-curve}.  The
eigenvalue corresponding to the minimum of the resonance tongue under
consideration is then determined, with the corresponding eigenvector
then giving the coefficients $a_j$ for the spectral representation of
$p_1(\tau)$, from which $q_1(\tau)$ can be calculated.  The adjoint
linear problem for $\widetilde{p}_1$ is solved in a similar way to
that for $p_1$.

Once a spectral representation for $q_1$ is found, $q_1^2$ is
calculated using a pseudospectral approach.  This then enables the
order $\epsilon^2$ problem for $h_2$, equation (\ref{eq:order2}), to
be written as a nonhomogeneous linear problem for $p_2$.  This can be
solved using the EISPACK routine {\tt rg}.  Finally, the coefficients
$\delta, \alpha$, $A$ and $B$ in the amplitude equation (\ref{landau})
are computed by calculating the various products of $p_1, p_2$ and
$q_1$ using a pseudospectral method and then calculating the inner
products.

Typically 20 Fourier modes sufficed to represent $p_1, p_2$ and $q_1$,
and 257 collocation points were adequate in the pseudospectral
calculation of the nonlinear terms.  Checks were done, with twice as
many modes and collocation points, to ensure that the results were
well converged.

In the next section we present plots of the ratio $A/B$ of the two
contributions to the cubic Landau coefficient for the case of
bifurcation to waves excited with forcing frequency ratio $2/3$, and
for the case of bifurcation to waves excited with forcing frequency
ratio $1/2$.

\section{Results.}
\label{sec:results}

We have calculated the coefficients $\delta, \alpha, A$ and $B$ in the
amplitude equation (\ref{landau}) for four cases.  These correspond
to the two lowest resonance tongues in the two cases $m=1,n=2$ and
$m=2,n=3$: whether harmonic or subharmonic waves bifurcate at a
smaller value of the driving amplitude depends on whether $\omega$ and
$\chi$ take values which are above or below the bicritical lines given
in Figure~\ref{bi-critical}. The other parameters used are the same as
those listed in the caption to Figure~\ref{bi-critical}.

\begin{figure}
\vskip 1.2truein
\centerline{
\epsfxsize=250pt
\epsffile{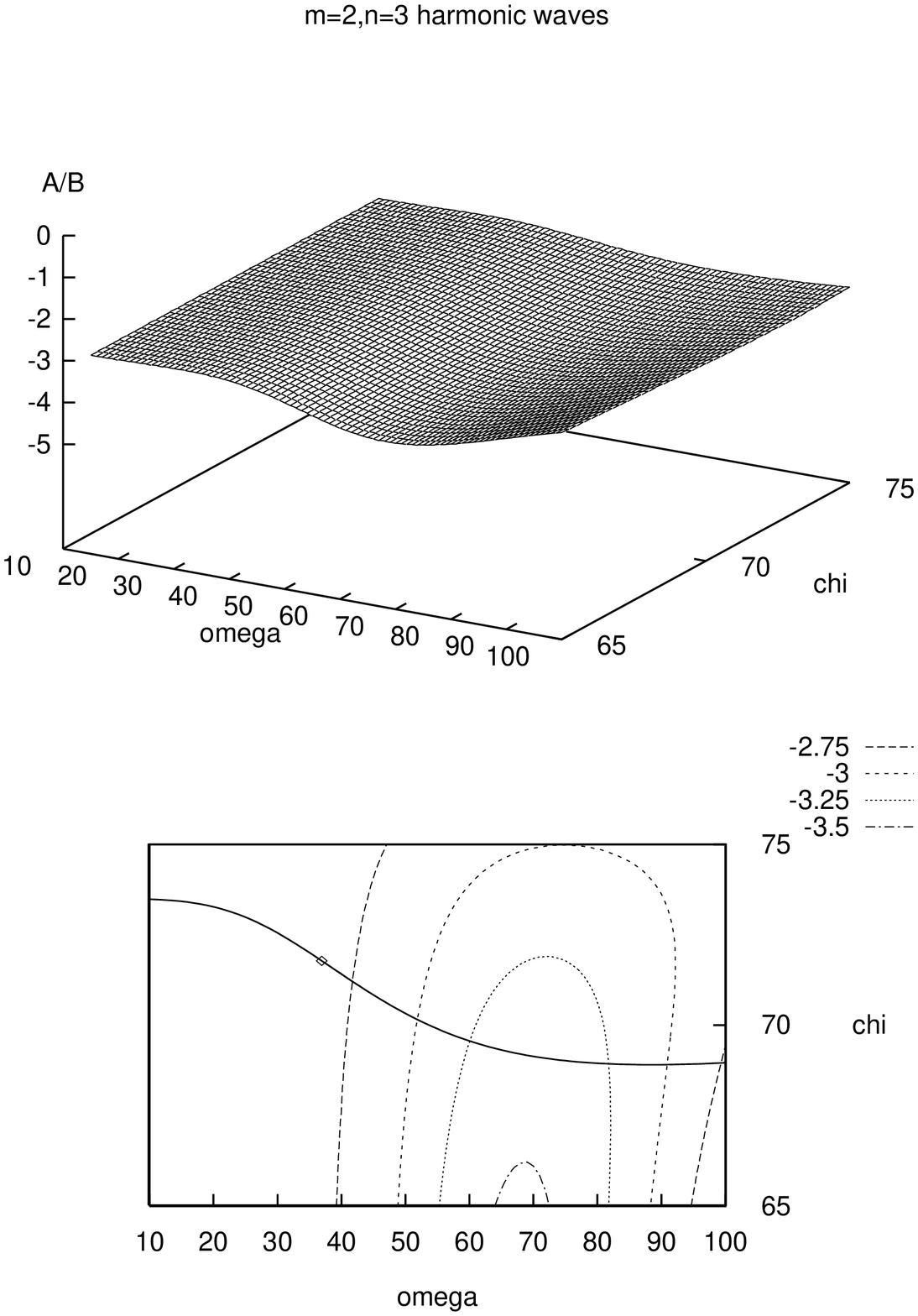}
\hskip -0.4truein
\epsfxsize=250pt
\epsffile{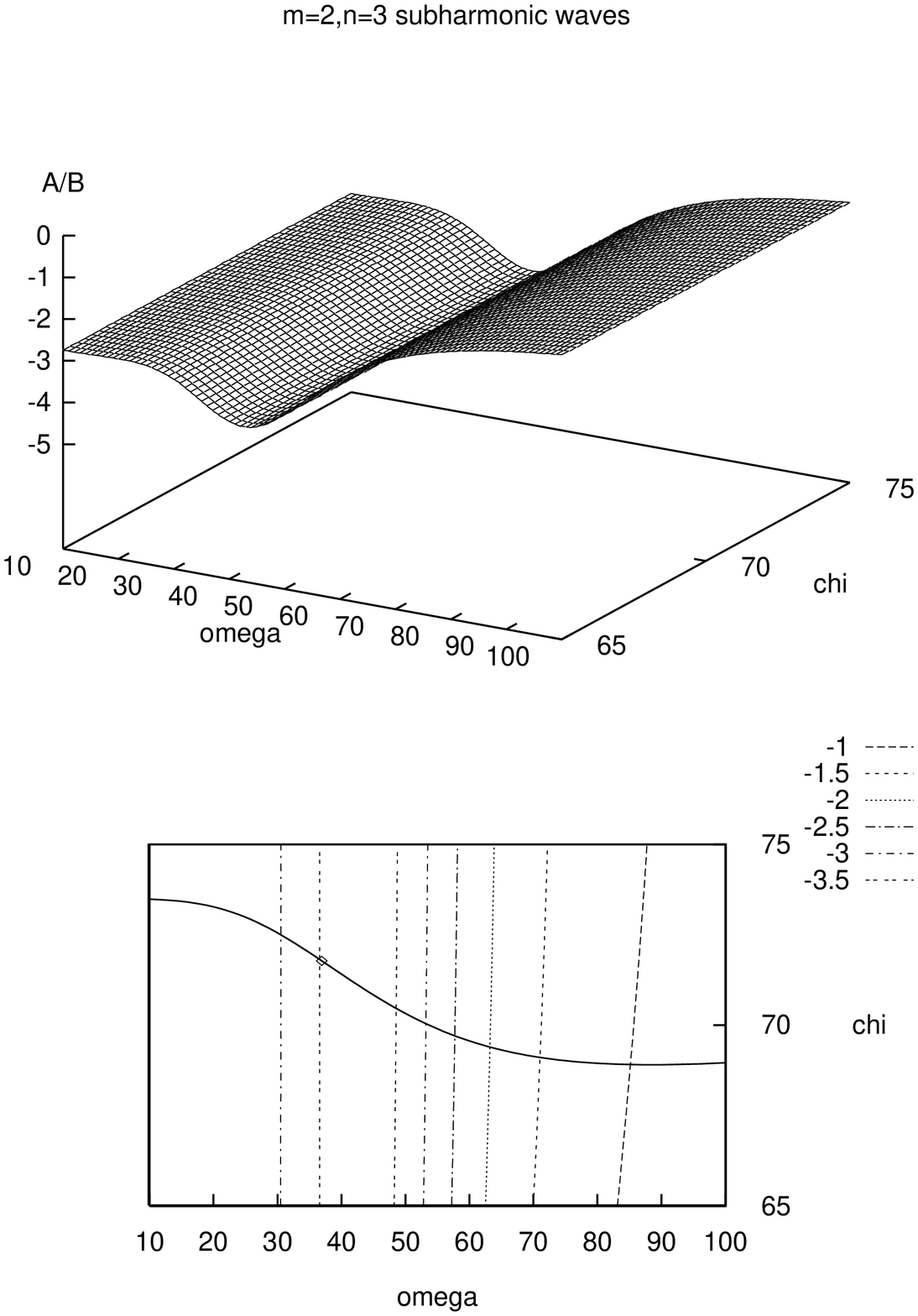}}
\vskip -0.3truein
\caption{Plots of $A/B$ 
as a function of $\chi$ and $\omega$, for $m=2$, $n=3$ and $\phi=0$ in
(\protect\ref{forcing}). Computations are done at the bifurcation point
to (sub)harmonic waves on the left (right). The fluid parameters are
$\Gamma=20.6\ dyn/cm$, $\nu=0.209\ cm^2/s$ and $\rho=0.95\ g/cm^3$.
We give both surface plots (top) and the corresponding contour plots
(below). In the contour plots we reproduce, from
Figure~\protect\ref{bi-critical}, the bicritical line in the
$\omega,\chi$--plane. The square on this curve gives the value of
$\omega$ where the 1/2 spatial resonance occurs.}
\label{fig:waves23}
\end{figure}

\begin{figure}
\vskip 1.2truein
\centerline{
\epsfxsize=250pt
\epsffile{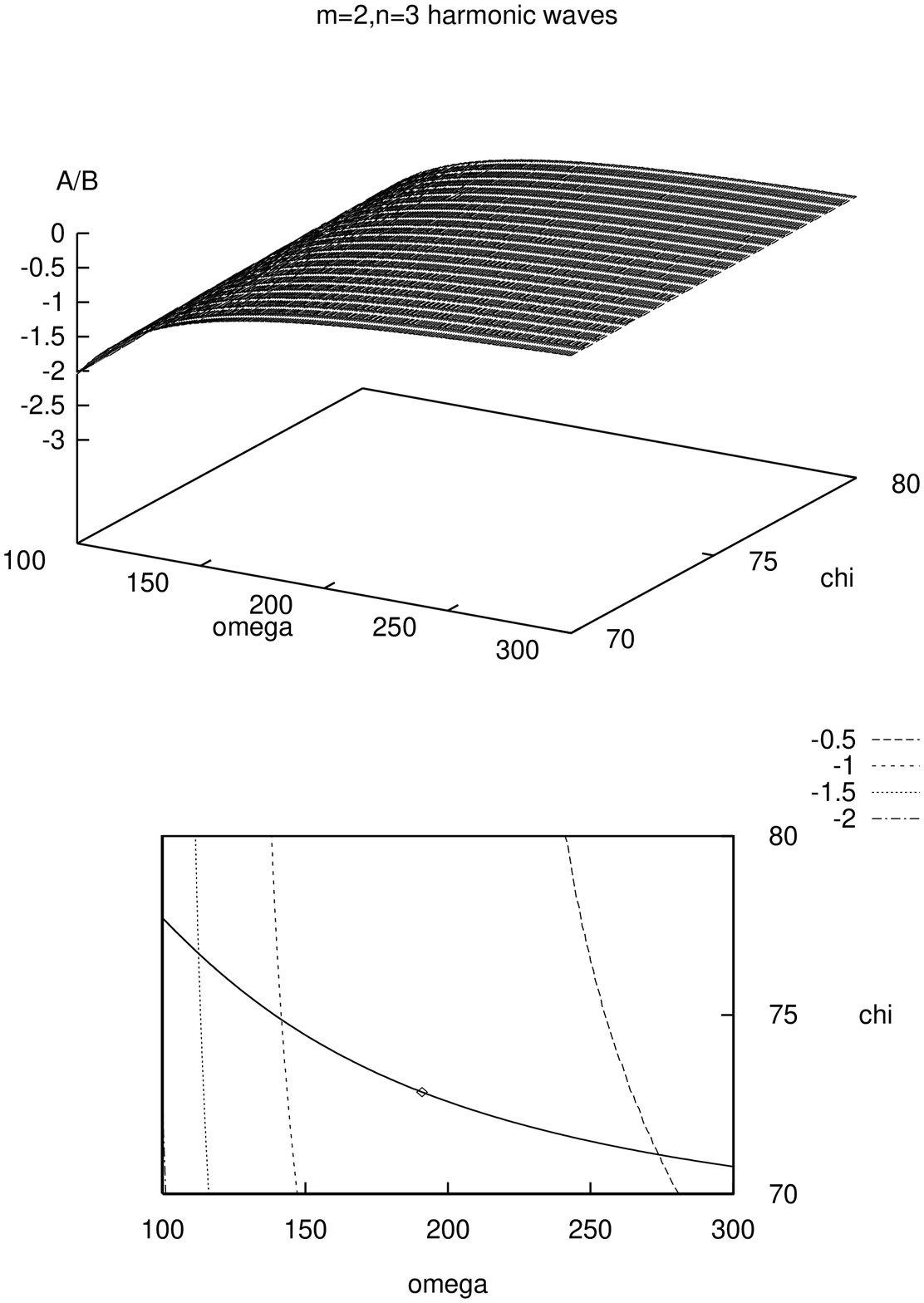}
\hskip -0.4truein
\epsfxsize=250pt
\epsffile{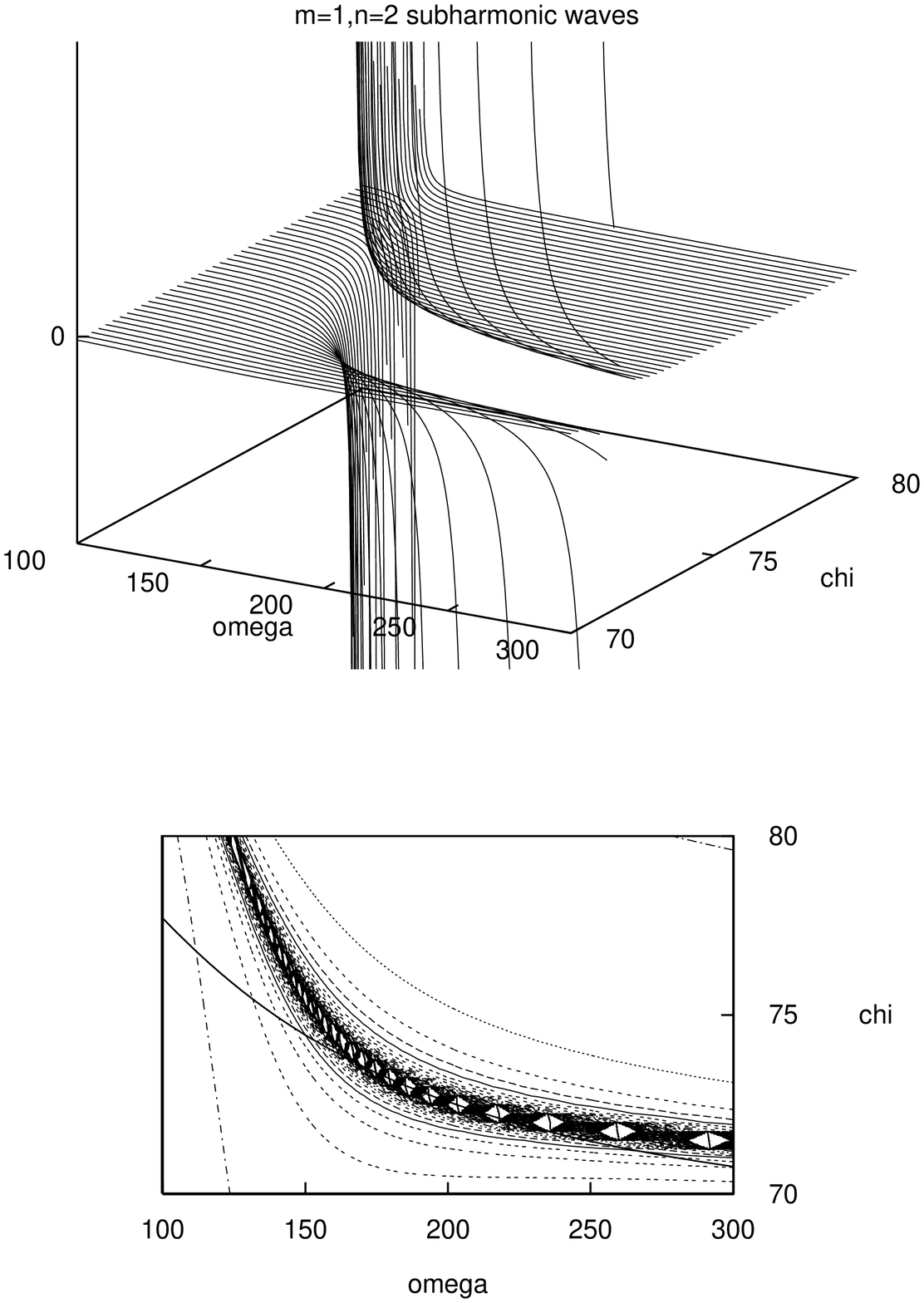}}
\vskip -0.3truein
\caption{
Plots of $A/B$ similar to those in Figure~\protect\ref{fig:waves23},
but computed for $m=1$ and $n=2$.  The divergence of $A/B$ for
subharmonic waves is discussed in the text. This divergence occurs on
a curve that is tangent to the bicritical line at the point
$\omega^{res}=191\ s^{-1},
\chi_c=72.83^\circ$; this is the point where there is a 1/2 spatial resonance
on the bicritical line.}
\label{fig:waves12}
\end{figure}

Figure~\ref{fig:waves23} shows the ratio $A/B$ of the two
contributions to the cubic Landau coefficient for the excitation
frequency ratio $2/3$ for the two tongues which bifurcate at lowest
amplitudes of the parametric excitation, $g_z$.  In each case the
upper graphs show the surface representing $A/B$ as a function of
$\omega$ and $\chi$.  The lower graphs show the corresponding contour
plots.  Superimposed on the contour plot is the bicritical line from
Figure~\ref{bi-critical} above: the square mark on this graph
indicates the point on the bicritical line at which the ratio between
the wavenumbers for the two minima is 2.  Note that in this case,
where the excitation frequency ratio is $2/3$, the bifurcation to the
harmonic waves occurs at lower amplitudes of the parametric excitation
$g_z$ for values of $\omega$ and $\chi$ below the bicritical line,
while the bifurcation to subharmonic waves occurs first for values of
$\omega$ and $\chi$ above the bicritical line.  The most striking
feature of the plots associated with the harmonic waves in
Figure~\ref{fig:waves23} is the insensitivity of the quantity $A/B$ to
the point on the bicritical line at which the ratio between the
wavenumbers is 2. Spatial resonance is also not important for the
subharmonic waves since the subharmonic tongue occurs at the higher
wavenumber in this case of even/odd excitation frequencies. This is in
marked contrast to the case of odd/even excitation, as demonstrated by
the graphs shown in Figure~\ref{fig:waves12}, which were computed for
an excitation frequency ratio $1/2$.  This time, it is the subharmonic
instability which occurs first for values of $\omega$ and $\chi$ below
the bicritical line and the harmonic instability which sets in first
for values of $\omega$ and $\chi$ above the bicritical line. As for
the previous graph, $A/B$ for the harmonic waves is unaffected by the
ratio between the wavenumbers and the parameter proximity to the
bicritical line.  However, in the case of subharmonic waves there is a
singularity occurring on the bicritical line at the point of spatial
resonance.  This is consistent with the discussion in Section~\ref{sec-res-one}
above, showing how critical the nature of the instability is in
determining whether or not spatial resonance is important to the
pattern formation process near the bicritical line.

\begin{figure}
\vskip 5truein
\centerline{
\hskip -0.5truein
\epsfxsize=300pt
\epsffile{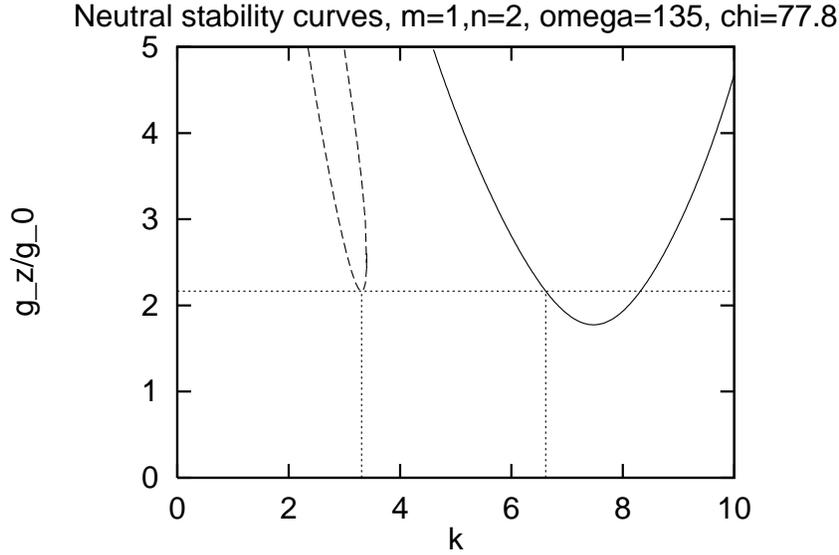}}
\vskip -4.5truein
\caption{Neutral stability curves for $\omega=135\ s^{-1}, \chi=77.8^\circ$.  The
harmonic (subharmonic) tongue is indicated by a solid (dashed) line.
The minimum of the subharmonic tongue occurs at $k_m=3.308\ cm^{-1}$,
when $g_z=2.1658\ g_0$. Also marked is the point where $2k_m$
intersects the harmonic tongue. Note that this intersection occurs at
the critical value of $g_z$, where bifurcation to subharmonic waves
occurs.}
\label{fig:neutral12}
\end{figure}

Note that in Figure~\ref{fig:waves12} there is in fact a line of
singularities tangent to the bicritical line at the point on the
bicritical line where the ratio between the wavenumbers is 1/2.  This
line of singularities can also be understood through spatial
resonance. Recall that above the bicritical line the subharmonic
instability occurs at a higher value of the excitation amplitude than
the harmonic instability, as shown in Figure~\ref{fig:neutral12} for
the values $\omega=135\ s^{-1}, \chi=77.8^\circ$.  These values were
chosen as representative of a point on the line of singularities shown
in Figure~\ref{fig:waves12}.  In this case the minimum of the
subharmonic tongue occurs at an excitation amplitude of $g_z=2.1658\
g_0$ and at a wavenumber of $k_m=3.308\ cm^{-1}$.  However, at this
excitation amplitude there is a bifurcation to harmonic waves occurring
at the wavenumber of $6.616\ cm^{-1}=2k_m$, and therefore spatial
resonance will occur.  Similarly, for values of $\omega$ and $\chi$ to
the right of the singularity which occurs on the bicritical line,
spatial resonance can occur between the minimum of the subharmonic
curve and the right side of the harmonic tongue.  Note that, in
practice, it is only close to the bicritical line that one would
expect to observe the effect of the spatial resonance on the
subharmonic waves, as in all other cases bifurcation to harmonic waves
occurs at significantly lower amplitudes of the excitation.  We also
note that for parameters {\it on} the bicritical line the center
manifold reduction described in Section~\ref{sec-res-one} can break
down.

\section{Conclusions.}
\label{sec:conclude}

In this paper we have investigated how normal form symmetries affect
the role of resonant triads in the pattern formation problem for
surface waves parametrically excited by two-frequency forcing.  We
focused on the behavior of the system near the bicritical point in
parameter space, where modes of wavenumber $k_m$ and $k_n$ lose
stability simultaneously with one mode associated with a Floquet
multiplier of $+1$ and the other associated with a Floquet multiplier
of $-1$. Our analysis shows that when the Floquet multiplier $+1$ is
associated with the smaller wavenumber, then quadratic terms may be
removed from the relevant amplitude equations.  In this case, the
contribution of resonant triads to the bifurcation problem is not
affected by proximity to the bicritical point. In contrast, when the
Floquet multiplier $-1$ is associated with the smaller wavenumber,
then the quadratic terms cannot be removed by a normal form
transformation. Hence, in this instance, resonant triads influence
greatly the bifurcation problem near the bicritical point.  The former
situation applies when the forcing frequency ratio is $m/n<1$, with
$m$ even and $n$ odd, while the latter situation occurs when $m$ is
odd and $n$ is even. Thus we expect normal form symmetries are
important to understanding the experimentally observed quasipatterns
and superlattice patterns \cite{ef:94,gollub}, which employ even/odd
forcing. Such effects are necessarily neglected in theoretical models,
such as the model of Lifshitz and Petrich~\cite{ref:lp97}, in which
quasipatterns form through a steady state bifurcation.

This paper demonstrates the influence of the normal form symmetries on
the bifurcation problem by considering the example of one-dimensional
waves, parametrically excited by two-frequency forcing. In this
one-dimensional problem a spatial resonance, involving Fourier modes
of wavenumber $k_m$ and $2k_m$, takes the place of a resonant
triad. Specifically, we focused on the situation where the bicritical
point involves modes $k_m$ and $k_n=2k_m$. We considered instability
to harmonic waves and subharmonic waves when the two-frequency forcing
was in ratio $m/n=1/2$ and $m/n=2/3$. Rather than perform the weakly
nonlinear analysis on the full hydrodynamic equations, we used the
simpler Zhang and Vi\~{n}als model~\cite{ref:zv97} that applies to a
deep layer of a nearly inviscid fluid. Consistent with our general
bifurcation theoretic analysis, we found that only in the case of
subharmonic waves, parametrically excited by $m/n=1/2$--forcing, did
the presence of the bicritical point lead to a diverging Landau
coefficient in the bifurcation problem. In the other cases, the Landau
coefficient was completely insensitive to any parameter proximity to
the bicritical point or to the spatial resonance.

Our work suggests that the spatially resonant triads, important to a
broad class of pattern formation problems, do not play an important
role in pattern formation of parametrically-excited surface waves near
the bicritical point in the case of $even/odd$ forcing. However, since
our analysis focused only on the case of a one-dimensional pattern
formation problem, more analysis in the two--dimensional case is
warranted. In the future we hope to carry out a more extensive
analysis of the contribution of resonant triads to two--dimensional
Faraday waves with $even/odd$ forcing. For instance, it would be of
interest to carry out such an analysis for the physical parameters of
the recent Faraday experiments in which superlattice patterns were
observed~\cite{gollub}. In these experiments the forcing frequencies
were in ratio $6/7$; the spatial Fourier transform of the harmonic
wave patterns exhibited peaks, some of which could be associated with
the two different frequency components of the forcing wave
form~\cite{gollub}.

Although the analysis presented in this paper was motivated by and
applied to the problem of Faraday waves with two-frequency forcing, we
expect many of the symmetry--based ideas to carry over to other
parametrically--excited pattern forming systems. We have in mind, for
example, the recent experiments on one-dimensional surface waves on
ferrofluids, which are excited by a time--periodic magnetic
field~\cite{rehberg}. In this system, both harmonic and subharmonic
response occur with single--frequency forcing.

\section*{Acknowledgements.}
We have benefited from discussions with Laurette Tuckerman. The research of
MS was supported by an NSF CAREER award DMS-9502266.

\end{document}